\magnification=1200
\vsize=7.5in
\hsize=5in
\tolerance 10000
\pageno=1

\def\up{{\vert \uparrow\rangle}}
\def\down{{\vert \downarrow \rangle}}
\baselineskip 12pt plus 2pt minus 2pt \centerline{\bf \quad ERRORS AND PARADOXES IN QUANTUM
MECHANICS}
\bigskip
According to one definition, a paradox is a statement that seems self-contradictory or
absurd but may be true; according to another, a paradox is a true self-contradiction and
therefore false. Let us define paradox to be an apparent contradiction that follows from
apparently acceptable assumptions via apparently valid deductions. Since logic admits no
contradictions, either the apparent contradiction is not a contradiction, or the
apparently acceptable assumptions are not acceptable, or the apparently valid deductions
are not valid.  A paradox can be useful in developing a physical theory; it can show that
something is wrong even when everything appears to be right.

Paradoxes in physics often arise as $\rightarrow$thought experiments.  For example, to
refute Aristotle's statement that a heavy body falls faster than a light one,
\hbox{Galileo [1]} invented a paradox: Suppose, with Aristotle, that a large stone falls
faster than a small stone. If the stones are tied together, the smaller stone will then
retard the large one. But the two stones tied together are heavier than either of them.
``Thus you see how, from your assumption that the heavier body moves more rapidly than the
lighter one, I infer that the heavier body moves more slowly."  Such free invention of
paradoxes as thought experiments marks especially the development of twentieth century
physics, i.e. of the relativity and quantum theories.

Both relativity theory and quantum theory are well supplied with paradoxes.  In
relativity theory, however, well known paradoxes such as the twin paradox have accepted
resolutions.  These paradoxes arise from intuitions, typically about simultaneity, that
relativity theory rendered obsolete.  By contrast, not all well known paradoxes of
quantum theory have accepted resolutions, even today. Below we briefly review seven
quantum paradoxes.

In keeping with our definition above, we do not distinguish between ``apparent" and
``true" paradoxes.  But we distinguish between apparent and true contradictions.  A true
contradiction is a fatal flaw showing that a physical theory is wrong.  By contrast,
apparent contradictions may arise from errors; they may also arise from a conceptual gap
in a theory, i.e. some ambiguity or incompleteness that is not fatal but can be removed by
further development of the theory.  Thus we can classify [2] physics paradoxes into three
classes:  Contradictions, Errors and Gaps.  The first three paradoxes below are examples
of a Contradiction, an Error and a Gap, respectively.
\bigskip

1.  By 1911, $\rightarrow$Rutherford and his co-workers had presented striking
experimental evidence (back-scattering of alpha particles) that neutral atoms of gold
have cores of concentrated positive charge.  According to classical electrodynamics, an
atom made of electrons surrounding a positive nucleus would immediately collapse; but
the gold foil in Rutherford's experiment evidently did not collapse.  This
contradiction between experimental evidence and classical theory was not merely
apparent:  it showed that atoms do not obey classical electrodynamics.  Faced with this
evidence, Bohr broke with classical theory and explained the stability of matter by
associating $\rightarrow$quantum numbers $n=1,2,3,\dots$ to the allowed orbits of
electrons in atoms.  Although $\rightarrow$Bohr's model described well only the
hydrogen atom, quantum numbers characterize all atoms.

2.  Einstein invented thought experiments to challenge Bohr's [3] principle of
$\rightarrow$complementarity.  One thought experiment involved two-slit interference.
(See Fig. 1.) Let a wave of (say) electrons of wavelength $\lambda$, collimated by a
screen with a single slit, impinge on a screen with two slits separated by $d$. An
electron interference pattern---dark lines with separation $D=\lambda L/d$---emerges on
a third screen a distance $L$ beyond the second.  In Fig. 1, however, the experiment is
modified to measure also the transverse recoil of the second screen (the screen with
the two slits).  Why the modification?  According to Bohr, a setup can demonstrate {\it
either} wave behavior (e.g. interference) of electrons {\it or} particle behavior (e.g.
passage through a single slit), but not {\it simultaneous} wave and particle behavior;
these two behaviors are complementary ($\rightarrow$``wave-particle duality") and no
setup can simultaneously reveal complementary behaviors.  Einstein's modified
experiment apparently shows electron interference while also revealing through which
slit each electron passes (e.g. an electron passing through the right slit makes the
screen recoil more strongly to the right) and thus contradicts the principle of
complementarity.

To analyze the modified experiment, let ${\bf p}^{(L)}$ and ${\bf p}^{(R)}$ denote the
momentum of an electron if it arrives at ${\cal P}$ via the left and right slits,
respectively, and let $p_\perp^{(L)}$ and $p_\perp^{(R)}$ denote the respective
transverse components.  From a measurement of the change in transverse momentum $p_s$
of the screen with accuracy $\Delta p_s \le$ $p_\perp^{(R)}-p_\perp^{(L)}$, we can
infer through which slit an electron passed.  But now apply $\rightarrow$Heisenberg's
uncertainty principle to the second screen:
$$
\Delta x_s \ge h /\Delta p_s \ge h / [p_\perp^{(R)}-p_\perp^{(L)}] ~~~,
$$
where $x_s$ is the transverse position of the second screen.  Similarity of triangles
in Fig. 1(b) implies that $\vert {\bf p}^{(R)} - {\bf p}^{(L)}\vert$ (which equals
$\vert p_\perp^{(R)} -p_\perp^{(L)} \vert$), divided by the electron's longitudinal
momentum $p_\parallel$, equals $d/L$.  The longitudinal momentum $p_\parallel$ is
$h/\lambda$ (assuming $p_\parallel$ large compared to the transverse momentum). Thus
$$
\Delta p_s < {d \over L} (h /\lambda ) ~~~~.
$$
We obtain $\Delta p_s < h/D$ and thus $\Delta x_s > D$. The uncertainty in the transverse
{\it position} $x_s$ of the screen, arising from an accurate enough measurement of its
transverse {\it momentum} $p_s$, is the distance $D$ between successive dark bands in the
interference pattern, and so the interference pattern is completely washed out.  Precisely
when Einstein's thought experiment succeeds in showing through which slit each electron
passes, it fails to show electron interference; that is, it {\it obeys} the principle of
complementarity after all.

3.  In 1931, Landau and Peierls [4] considered the following model measurement of the
electric field ${\bf E}$ in a region.  Send a charged test particle through the region;
the electric field deflects the particle, and the change in the momentum ${\bf p}$ of
the test particle is a measure of ${\bf E}$. But an accelerated, charged particle
radiates, losing an unknown fraction of its momentum to the electromagnetic field.
Reducing the charge on the test particle reduces radiation losses but then ${\bf p}$
changes more slowly and the measurement lasts longer (or is less accurate).  On the
basis of their model, Landau and Peierls concluded that an instantaneous, accurate
measurement of ${\bf E}$ is impossible.  They obtained a lower bound $\Delta \vert {\bf
E}\vert \ge \sqrt{\hbar c}/(cT)^2$ as the minimum uncertainty in a measurement of
$\vert {\bf E}\vert$ lasting a time $T$. Their conclusion is paradoxical because it
leaves the instantaneous electric field ${\bf E}$ with no theoretical or experimental
definition.  However, the Landau-Peierls model measurement is too restrictive. Bohr and
Rosenfeld [5] found it necessary to modify the model in many ways; one modification was
to replace the (point) test particles of Landau and Peierls with extended test bodies.
In their modified model, they showed how to measure electric (and magnetic) fields
instantaneously.  Note that the electric field is not a {\it canonical} variable, i.e.
it is not one of the generalized coordinates and momenta appearing in the associated
Hamiltonian.  (It depends on the time derivative of ${\bf A}$, the electromagnetic
vector potential, which {\it is} a canonical variable.)  The resolution of this sort of
paradox is that quantum measurements of canonical and noncanonical variables differ
systematically [6].

4. $\rightarrow${\it Zeno's paradoxes} are named for the Greek philosopher who tried to
understand motion over shorter and shorter time intervals and found himself proving
that motion is impossible. The {\it quantum} Zeno paradox [7] seems to prove that
quantum evolution is impossible. Consider the evolution of a simple quantum system:  a
spin-1/2 atom precesses in a constant magnetic field. If we neglect all but the spin
degree of freedom, represented by the $\rightarrow$Pauli spin matrices $\sigma_x$,
$\sigma_y$ and $\sigma_z$, the Hamiltonian is
$$
H = \mu B \sigma_z
$$
where the direction of the magnetic field defines the $z$-axis and $\mu$ is the Bohr
magneton.  Suppose that at time $t=0$ the state is
$$
\vert \psi (0) \rangle ={1\over\sqrt{2}} \left[ \up + \down \right]
$$
(where $\sigma_z\up =\up$ and $\sigma_z\down=-\down$). Solving
$\rightarrow$Schr\"odinger's equation
$$
i\hbar {d \over {dt}}  \vert \psi (t)\rangle =H \vert \psi\rangle~~~,
$$
we obtain the time evolution:
$$\eqalign{
\vert \psi (t) \rangle &= e^{-iHt/\hbar} \vert \psi (0)\rangle \cr &= {1\over\sqrt{2}}
\left[ e^{-i\mu Bt/\hbar} \up +e^{i\mu Bt/\hbar} \down \right]~~~~.\cr }
$$
At $t=0$, a measurement of $\sigma_x$ is sure to yield 1; at time $t=T\equiv h /4\mu
B$, the $\sigma_x$ measurement is sure to yield $-1$; at intermediate times, a
measurement may yield either result.

At no time does a measurement of $\sigma_x$ yield a value other than $1$ and $-1$; the
spin component $\sigma_x$ apparently $\rightarrow$jumps discontinuously from 1 to $-1$,
defining a moment in time by jumping. {\it When} does the spin jump? We cannot predict
when it will jump, but we can make many measurements of $\sigma_x$ between $t=0$ and
$t=T$. The jump in $\sigma_x$ must occur between two successive measurements. When it
does, we will know when the jump occurred, to an accuracy $\Delta t$ equal to the time
between the measurements. But now we apparently violate the uncertainty relation for
energy and time:
$$
\Delta E \Delta t \ge \hbar /2~~~~.
$$
Here $E$ is the energy of the measured system and $t$ is time as defined {\it by} the
system.  (Although $t$ is not an operator, we can define $t$ via an operator that
changes smoothly in time, and then derive $\Delta E \Delta t \ge \hbar /2$ indirectly
[8].) The problem is that the uncertainty $\Delta E$ in the energy cannot be greater
than the difference $2\mu B$ between the two eigenvalues of $H$; but the measurements
can be arbitrarily dense, i.e. $\Delta t$ can be arbitrarily small.

Since quantum mechanics will not allow a violation of the uncertainty principle, we may
guess that the atomic spin will simply refuse to jump! A short calculation verifies
this guess.  Consider $N$ measurements of $\sigma_x$, at equal time intervals, over a
period of time $T$.  The interval between measurements is $T/N$. What is the
probability of finding the spin unchanged after the first measurement?  The state at
time $t=T/N$ is
$$
{1\over\sqrt{2}} \left[ e^{-i\mu BT/N\hbar} \up +e^{i\mu BT/N\hbar} \down \right]~~~,
$$
so the probability of finding the spin unchanged is $\cos^2 (\mu BT/N\hbar )$. Hence
the probability of finding the spin unchanged at time $T$, after $N$ measurements, is
$\cos^{2N} (\mu BT/N\hbar )$. As $N$ approaches infinity, $\cos^{2N} (\mu BT/N\hbar )$
approaches 1:  the spin never jumps.  Here quantum evolution is impossible.  But
consider a dual experiment: instead of $N$ measurements of $\sigma_x$ on an atom in a
magnetic field, consider $N$ measurements of $\sigma_x \cos (2\mu Bt/\hbar ) +\sigma_y
\sin (2\mu Bt/\hbar )$, at equal time intervals, on an atom in no magnetic field
($H=0$). In the limit $N \rightarrow \infty$, the atom precesses:  each measurement of
$\sigma_x \cos (2\mu Bt/\hbar) + \sigma_y \sin (2\mu Bt /\hbar)$ yields 1.  Experiments
from 1990 on have progressively demonstrated such quantum Zeno effects.

5. A thought experiment due to $\rightarrow$Einstein, Podolsky and Rosen [9] (EPR)
shows how to measure precisely the position ${\bf x}_A(T)$ {\it or} the momentum ${\bf
p}_A (T)$ of a particle A at a given time $T$, {\it indirectly} via a measurement on a
particle B that once interacted with A.  The measurement on B is spacelike separated
from ${\bf x}_A(T)$, and so it cannot have any measurable effect on ${\bf x}_A(T)$ or
${\bf p}_A(T)$ (no superluminal signalling).  It is indeed reasonable to assume
($\rightarrow$ ``Einstein locality") that the measurement on B has no effect whatsoever
on ${\bf x}_A(T)$ or ${\bf p}_A(T)$; thus ${\bf x}_A(T)$ and ${\bf p}_A(T)$ are
simultaneously defined (in the sense that either is measurable without any effect on
the other) and a particle has a precise position and momentum simultaneously. Since
quantum mechanics does not define the precise position and momentum of a particle
simultaneously, quantum mechanics does not completely describe particles. EPR
envisioned a theory that would be {\it consistent} with quantum mechanics but more
complete, just as statistical mechanics is consistent with thermodynamics but more
complete.

Almost 30 years after the EPR paper, $\rightarrow$Bell [10] proved a startling,
and---to Bell himself---disappointing theorem: Any more complete theory of the sort
envisioned by EPR would contradict quantum mechanics!  Namely, the correlations of any
such theory must obey $\rightarrow$Bell'a inequality; but according to quantum
mechanics, some correlations of $\rightarrow$entangled states of particles A and B
violate Bell's inequality.  If quantum mechanics is correct, then there can be no
theory of the sort envisioned by EPR.  $\rightarrow$Experiments have, with increasing
precision and rigor, demonstrated violations of Bell's inequality and ruled out any
theory of the sort envisioned by EPR.

6.  In 1927, at the fifth Solvay congress, Einstein presented ``a very simple
objection" to the probability interpretation of quantum mechanics.  According to
quantum mechanics, the state of an electron approaching a photographic plate is an
extended object; the probability density for the electron to hit varies smoothly over
the plate. Once the electron hits somewhere on the plate, however, the probability for
the electron to hit anywhere else drops to zero, and the state of the electron
$\rightarrow$collapses instantaneously.  But instantaneous collapse of an extended
object is not compatible with relativity.  A related paradox is the following.
\hbox{Fig. 2} shows two atoms, prepared in an entangled state at {\it O}, flying off in
different directions. (For simplicity, assume that they separate at nonrelativistic
speeds.) One atom enters the laboratory of Alice, who measures a component of its spin
at $a$; the other enters the laboratory of Bob, who measures a component of its spin at
$b$. After Alice's measurement, the atoms are not in an entangled state anymore, hence
collapse cannot occur anywhere outside the past light cone of $a$. Likewise, collapse
cannot occur anywhere outside the past light cone of $b$.  Hence collapse cannot occur
anywhere outside the {\it intersection} of the past light cones of $a$ and $b$.  But
then, in the inertial reference frame of Fig. 2, the state of the atoms just before
either measurement is a product (collapsed) state, not an entangled state.  Now this
conclusion contradicts the fact that, by repeating this experiment on many pairs of
atoms, Alice and Bob can obtain violations of Bell's inequality, i.e. can demonstrate
that the atomic spins were in an entangled state until Bob's measurement. This paradox
shows that there can be no Lorentz-invariant account of the collapse.  In general,
observers in different inertial reference frames will disagree about collapse. They
will not disagree about the results of local measurements, because local measurements
are spacetime events, hence Lorentz invariant; but they will have different accounts of
the collapse of nonlocal states. Collapse is Lorentz {\it co}variant [11].

7.  $\rightarrow$Schr\"odinger's Cat is a paradox of quantum evolution and measurement.
For simplicity, let us consider just the $\sigma_z$ degree of freedom of spin-1/2 atoms
and define a superposition of the two normalized eigenstates $\up$ and $\down$ of
$\sigma_z$:
$$
\vert \Psi_{\alpha \beta}\rangle =\alpha \up + \beta \down~~~;
$$
we assume $\vert \alpha \vert^2 +\vert \beta\vert^2 =1$.  The Born probability rule
states that a measurement of $\sigma_z$ on many atoms prepared in the state $\vert
\Psi_{\alpha \beta}\rangle$ will yield a fraction approaching $\vert \alpha\vert^2$ of
atoms in the state $\up$ and a fraction approaching $\vert \beta \vert^2$ of atoms in
the state $\down$. If quantum mechanics is a complete theory, it should be possible to
describe these measurements themselves using Schr\"odinger's equation. We can describe
a measurement on an atom abstractly by letting $\vert \Phi_0 \rangle$ represent the
initial state of a measuring device, and letting $\vert \Phi_\uparrow \rangle$ or
$\vert \Phi_\downarrow\rangle$ represent the final state of the measuring device if the
state of the atom was $\up$ or $\down$, respectively.  If the Hamiltonian for the
measuring device and atom together is $H$, during a time interval $0 \le t\le T$ that
includes the measurement, then the Schr\"odinger equation implies
$$
\eqalign{ e^{-i\int_0^T Hdt/\hbar} \up \otimes \vert \Phi_0 \rangle&=\up \otimes \vert
\Phi_\uparrow \rangle~~~, \cr e^{-i\int_0^T Hdt/\hbar} \down \otimes \vert \Phi_0
\rangle&=\down \otimes \vert \Phi_\downarrow \rangle~~~~.\cr }
$$
(The spin states do not change as they are eigenstates of the measured observable
$\sigma_z$.) If the initial spin state is neither $\up$ nor $\down$ but the
superposition $\vert \Psi_{\alpha \beta} \rangle$, the evolution of the superposition
is the superposition of the evolutions:
$$
e^{-i\int_0^T Hdt/\hbar} \vert \Phi_{\alpha \beta } \rangle \otimes \vert \Phi_0
\rangle =\alpha \up \otimes \vert \Phi_\uparrow \rangle +\beta \down \otimes \vert
\Phi_\downarrow \rangle~~~~.
$$
The right side of this equation, however, does not describe a completed measurement at
all:  the measuring device remains entangled with the atom in a superposition of
incompatible measurement results.  It does not help to couple additional measuring
devices to this device or to the atom; since the Schr\"odinger equation dictates
linear, unitary evolution, additional devices will simply participate in the
superposition rather than collapse it.  Even a cat coupled to the measurement will
participate in the superposition.  Suppose the measuring device is triggered to release
poison gas into a chamber containing a cat, {\it only} if the spin state of the
measured atom is $\up$. The state of the atom, measuring device and cat at time $t=T$
will be a superposition of $\up \otimes \vert \Phi_\uparrow \rangle \otimes \vert
\rm{dead}\rangle$ and  $\down \otimes \vert \Phi_\downarrow \rangle \otimes \vert
\rm{live}\rangle$ with coefficients $\alpha$ and $\beta$, respectively.  So we do not
know how to describe even one measurement using Schr\"odinger's equation.
\bigskip

Paradoxes 1-4 and 6 and their resolutions are not controversial.  Paradoxes 5 and 7,
however, do excite controversy.  For many physicists, the EPR paradox and Bell's
theorem remain unresolved because, for them, renouncing the ``reasonable" assumption of
EPR is just not a resolution.  As one distinguished physicist put it [12], ``Anybody
who's not bothered by Bell's theorem has to have rocks in his head."  (No such
statement would apply to any well known paradox in relativity theory.)

The Schr\"odinger Cat paradox has been resolved several times over---with spontaneous
$\rightarrow$``collapse" of quantum states [13], nonlocal $\rightarrow$``hidden
variables" [14], $\rightarrow$``many (parallel) worlds" [15] and future boundary
conditions [16] (conditions on the future state in a $\rightarrow$``two-state"
formalism [17])---but since experiments are consistent with all these resolutions,
there is no one accepted resolution, at least within nonrelativistic quantum mechanics.
The predictions of quantum mechanics with and without collapse differ, but the
differences are (so far) not accessible to experiment. (There is even a proof [18] that
if quantum mechanics is correct and an experiment could verify that a cat is in the
superposition $\alpha \vert \rm{dead} \rangle +\beta\vert \rm{live} \rangle$, i.e. if
it could verify that collapse has not occurred, the same experiment could transform the
state $\vert \rm{dead} \rangle$ into the state $\vert \rm{live} \rangle$, i.e. it could
revive a dead cat.)  However, it is doubtful whether collapse or hidden-variable
theories can be made relativistic; hence resolutions via many worlds or future boundary
conditions, which require neither collapse nor ``hidden" superluminal signalling, seem
preferable so far.

\goodbreak

\bigskip

\centerline{\bf \quad References}
\medskip
\noindent [Primary]

[1] {Galileo Galilei, {\it Dialogues Concerning Two New Sciences}, trans. H. Crew and A.
de Salvio (Dover, New York, 1954), p. 63.}

[2] {Y. Aharonov and D. Rohrlich, {\it Quantum Paradoxes: Quantum Theory for the
Perplexed} (Wiley-VCH, Weinheim, 2005), Chap. 1.}

[3] {N. Bohr, ``Discussion with Einstein on epistemological problems in atomic physics",
in {\it Albert Einstein: Philosopher--Scientist}, ed. Paul A. Schilpp (Tudor Pub. Co., New
York, 1951), pp. 201-41.}

[4] {L. Landau and R. Peierls, ``Erweiterung des Unbestimmtheitsprinzips f\"ur die
relativistische Quantentheorie", {\it Z. Phys.} {\bf 69}, 56-69 (1931); trans. ``Extension
of the uncertainty principle to the relativistic quantum theory", in {\it Collected Papers
of Landau}, ed. D. ter Haar (Gordon and Breach New York, 1965), pp. 40-51; also in J. A.
Wheeler and W. H. Zurek, {\it Quantum Theory and Measurement} (Princeton U. Press,
Princeton, 1983), pp. 465-476.}

[5] {N. Bohr and L. Rosenfeld, ``Zur Frage der Messbarkeit der elektromagnetischen
Feldgr\"ossen", {\it Det Kgl. Danske Vid. Selsk. Mat.--fys. Medd.} {\bf XII}, no. 8, 65
pp. (1933); trans. by A. Petersen, ``On the question of the measurability of
electromagnetic field quantities", in {\it Selected Papers of L\'eon Rosenfeld}, eds. R.
S. Cohen and J. Stachel (Reidel, Dordrecht, 1979), pp. 357-400; also in J. A. Wheeler and
W. H. Zurek, {\it op. cit.} pp. 479-522.}

[6] {Y. Aharonov and J. L. Safko, ``Measurement of noncanonical variables", {\it Ann.
Phys.} {\bf 91}, 279-294 (1975); see also Y. Aharonov and D. Rohrlich {\it op. cit.},
Chap. 8.}

[7] {B. Misra and E. C. G. Sudarshan, ``The Zeno's paradox in quantum theory", {\it J.
Math. Phys.} {\bf 18}, 756-763 (1977).}

[8] Aharonov and Rohrlich, {\it op. cit.}, Sect. 8.5.

[9] {A. Einstein, B. Podolsky and N. Rosen, ``Can quantum-mechanical description of
physical reality be considered complete?" {\it Phys. Rev.} {\bf 47}, 777-780 (1935).}

[10] {J. S. Bell, ``On the Einstein--Podolsky--Rosen paradox", {\it Physics} {\bf 1},
195-200 (1964).}

[11] {Y. Aharonov and D. Z. Albert,  ``Is the usual notion of time evolution adequate
for quantum-mechanical systems? II. Relativistic considerations", {\it Phys. Rev.} {\bf
D29}, 228-234 (1984); Aharonov and Rohrlich {\it op. cit.}, Chap. 14.}

[12]  Quoted in N. D. Mermin, ``Is the moon there when nobody looks? Reality and the
quantum theory", {\it Physics Today} {\bf 38}, no. 4 (April), pp. 38-47 (1985).

[13] G. C. Ghirardi, A. Rimini and T. Weber, ``Unified dynamics for microscopic and
macroscopic systems", {\it Phys. Rev.} {\bf D34}, 470-491 (1986); ``Disentanglement of
quantum wave functions: answer to `Comment on ``Unified dynamics for microscopic and
macroscopic systems'''", {\it Phys. Rev.} {\bf D36}, 3287-3289 (1987); P. Pearle,
``Combining stochastic dynamical state-vector reduction with spontaneous localization",
{\it Phys. Rev.} {\bf A39}, 2277-2289 (1989); G. C. Ghirardi, P. Pearle and A. Rimini,
``Markov processes in Hilbert space and continuous spontaneous localization of systems of
identical particles" {\it Phys. Rev.} {\bf A42}, 78-89 (1990); P. Pearle, ``Relativistic
collapse model with tachyonic features", {\it Phys. Rev.} {\bf A59}, 80-101 (1999).

[14] D. Bohm, ``A suggested interpretation of the quantum theory in terms of `hidden'
variables", parts I and II, {\it Phys. Rev.} {\bf 85} 166-179 and 180-193 (1952),
reprinted in Wheeler and Zurek, {\it op. cit.}, pp. 369-82 and 383-96.

[15] H. Everett, III, ```Relative state' formulation of quantum mechanics", {\it Rev. Mod.
Phys.} {\bf 29}, 454-462 (1957), reprinted in Wheeler and Zurek, {\it op. cit.}, pp.
315-23; J. A. Wheeler, ``Assessment of Everett's `relative state' formulation of quantum
theory", {\it Rev. Mod. Phys.} {\bf 29}, 463-465 (1957).

[16] Y. Aharonov and E. Gruss, eprint quant-ph/0507269 (2005); Aharonov and Rohrlich, {\it
op. cit.}, Sect. 18.3.

[17] Y. Aharonov and L. Vaidman, ``On the two-state vector reformulation of quantum
mechanics", {\it Phys. Scrip.} {\bf T76}, 85-92 (1998).

[18] {Aharonov and Rohrlich {\it op. cit.}, p. 131.}

\bigskip

\centerline{\bf \quad Figure Captions}
\medskip
Fig. 1. (a) A two-slit interference experiment adapted for measuring the transverse
momentum of the middle screen.  (b) The second and third screens seen from above, with
interfering electron paths and corresponding momenta.

\medskip
Fig. 2.  Two atoms, produced in an entangled state at {\it O}, fly off in opposite
directions (solid lines) in this spacetime figure.  Alice measures a spin component of
one atom at $a$; Bob measures a spin component of the other atom at $b$. Collapse
cannot occur anywhere outside the past light cones of $a$ and $b$ (dotted lines), hence
it cannot occur anywhere outside the {intersection} of their past light cones (shaded
region).

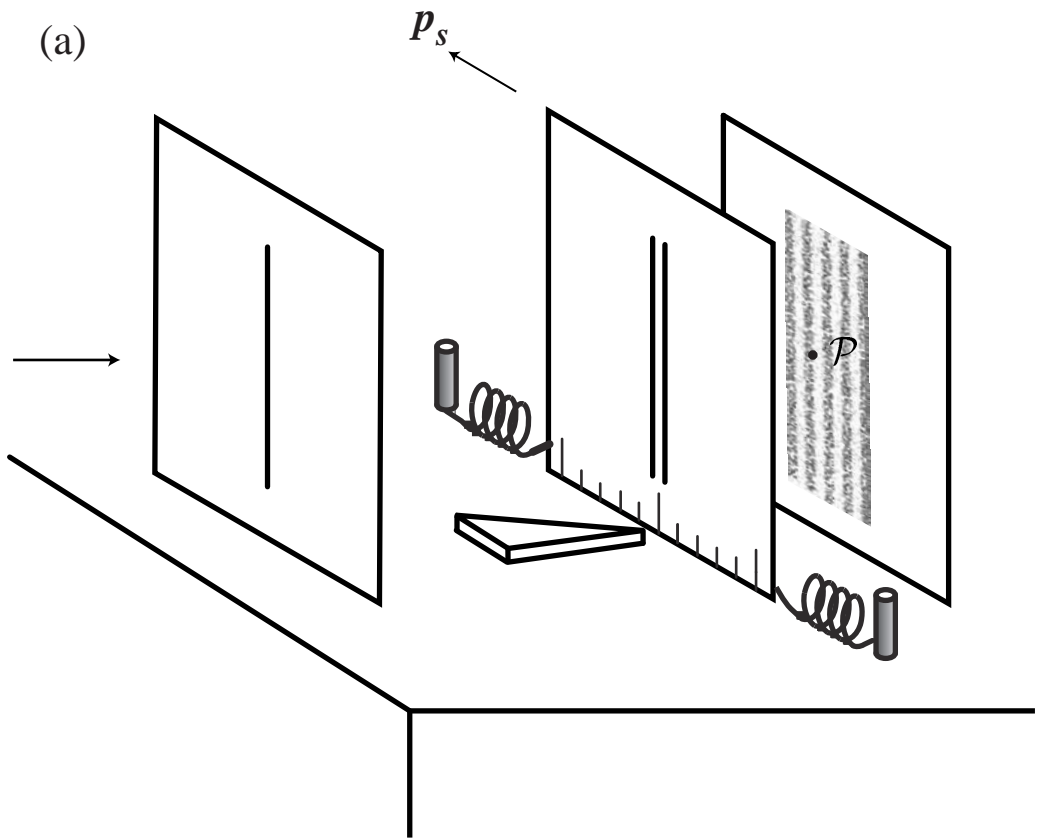
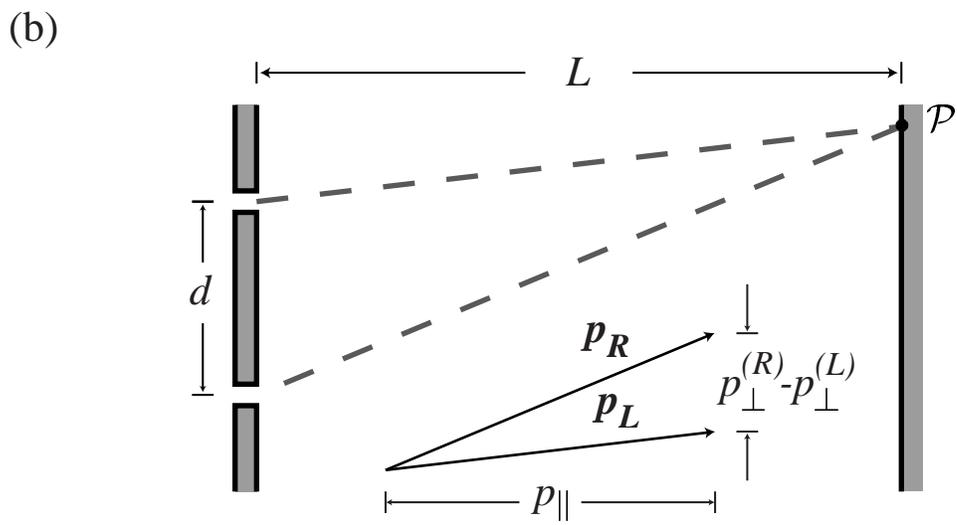
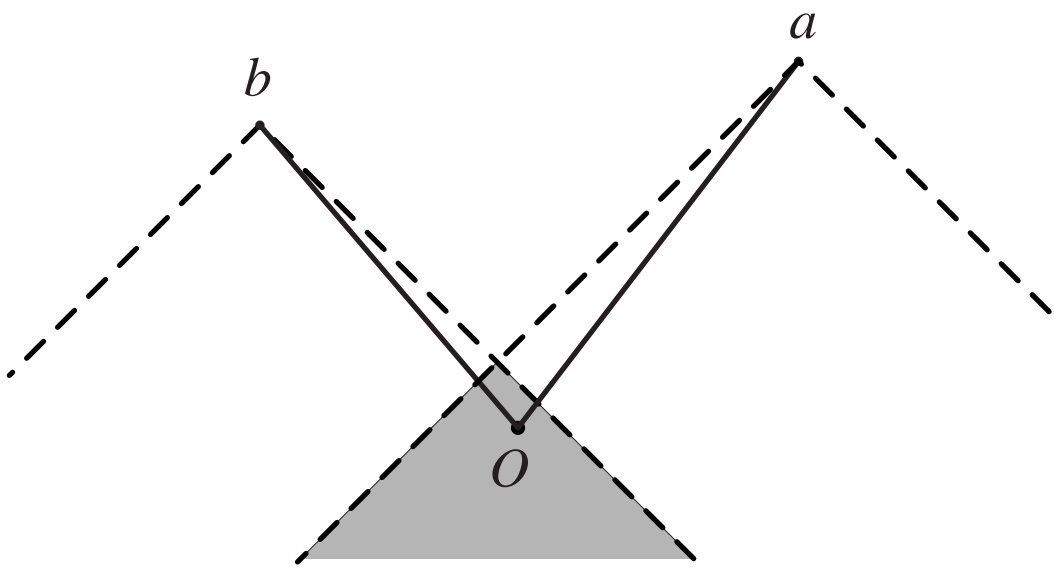
\end